\newcommand{\les}{\stackrel{<}{{}_{\sim}}}

\newcommand{\VEV}[3]{\left\langle #1\left| #2 \right| #3\right\rangle}

\documentclass[
    ,final            
  ]
  {aipproc}

\layoutstyle{6x9}

\begin{document}

\title{Calculation of CP Violation in Non-leptonic Kaon Decay on the 
Lattice }

\author{J.~Noaki for RBC Collaboration}{
  address={RIKEN BNL Research Center, Bldg. 510A,
  Brookhaven National Laboratory, Upton NY, 11973}
}

\begin{abstract}
We give a progress report of our lattice calculation of direct and 
indirect CP violation in kaon decays, parametrized as $\epsilon'/\epsilon$ 
and $B_K$, which require non-perturbative calculation of the matrix 
elements of the Standard Model effective Hamiltonian.
\end{abstract}

\maketitle

\vspace*{-1.2cm}

\section{INTRODUCTION}
In the investigations of the Standard Model, a very important issue is the
theoretical treatment of $K\to\pi\pi$ decay to the accuracy such that 
comparison with the experimental results is possible. In particular, 
the ratio of direct and indirect CP violation, $\epsilon'/\epsilon$, 
has been determined experimentally \cite{EXPERIMENT} in 
recent years and theoretical calculation is desired to test 
Kobayashi-Maskawa theory. In the theoretical calculation, 
numerical simulation of lattice QCD is the most systematic method to
estimate the non-perturbative effect of QCD which is the main source of 
the error. Using the operator product expansion, the interaction 
in this decay is written as 
$H_W= \frac{G_FV_{us}V^*_{ud}}{\sqrt{2}}\sum_{i}W_i(\mu)Q_i, $
where the coefficients $W_i$ contain the effects of the energy scales higher
than the matching point $\mu$ and can be obtained perturbatively \cite{Buras}. 
Non-perturbative QCD effects will appear in $K\to\pi\pi$ matrix elements
of the local operators $\VEV{\pi\pi}{Q_i}{K}$, which should be
calculated on the lattice.
A couple of years ago,
CP-PACS and RBC Collaboration~\cite{CPPACSep, RBCep} calculated all 
of the matrix elements using the domain-wall fermion formalism
~\cite{Kaplan, FurmanShamir} to realize the chiral symmetry required 
in this calculation
and reported small and negative values of $\epsilon'/\epsilon$
in conflict with the experimental result. Another work using staggered fermion 
has obtained a larger negative value~\cite{PekurovskyKilcup}.
In these works, however, there are several uncontrolled systematic
errors such as 1) the effect of the small, but non-zero, chiral symmetry 
breaking, 2) the effect of finite lattice spacing,
3) the effect of 
the perturbative treatment of the charmed quark in the matrix elements, 
4) quenching effect, and 5) $K \to \pi\pi$ matrix elements are obtained 
from $K \to \pi$ and $K \to 0$ (vacuum) by using lowest order 
chiral perurbation theory~\cite{BernardSoni}.

In order to examine all of these systematic errors except the fifth
one, we are performing two types of numerical simulation with 
domain-wall fermion and the DBW2 gluonic action~\cite{RBCDBW2} to
improve the chiral symmetry on the lattice.
``Numerical Simulation I'' is the quenched calculation including 
directly the effect of the charm quark on the lattice.
The degree of chiral symmetry 
breaking is decreased by a factor 1/10 compared with the previous work 
of RBC Collaboration. In addtion, we are generating gauge
configurations with $N_f=2$ dynamical quarks~\cite{RBCdyn} in 
``Numerical Simulation II.''
In the rest of this article, we present the contents of these numerical 
simulations and report preliminary results of the matrix elements which 
numerically dominate $\epsilon'/\epsilon$ and kaon B-parameter $B_K$.
\vspace*{-0.2cm}

\section{NUMERICAL SIMULATION I }
We are generating gauge configurations on a relatively fine $24^3\times 48$ 
lattice with the scale  $a^{-1}=2.86(9)$.
The residual quark mass $m_{\rm res}$ which measures the chiral 
symmetry breaking is as small as $\les$ 0.3 MeV. 
Since quark mass $m_fa$ is introduced as a parameter 
of the boundary condition in the fifth dimension in domain-wall QCD,
the localization of chiral modes on domain-walls in the fifth dimension 
tends to fail for a heavy quark mass $m_f$.
However, our small lattice spacing made the value of $m_ca$ acceptable 
as a domain-wall fermion: $m_ca\approx$ 0.45. We found that, around this
value, the behavior of wave function in the fifth dimension is 
qualitatively same as the case of much smaller quark mass. 

\begin{figure}
\begin{minipage}{0.48\linewidth}
  \includegraphics[height=.28\textheight]{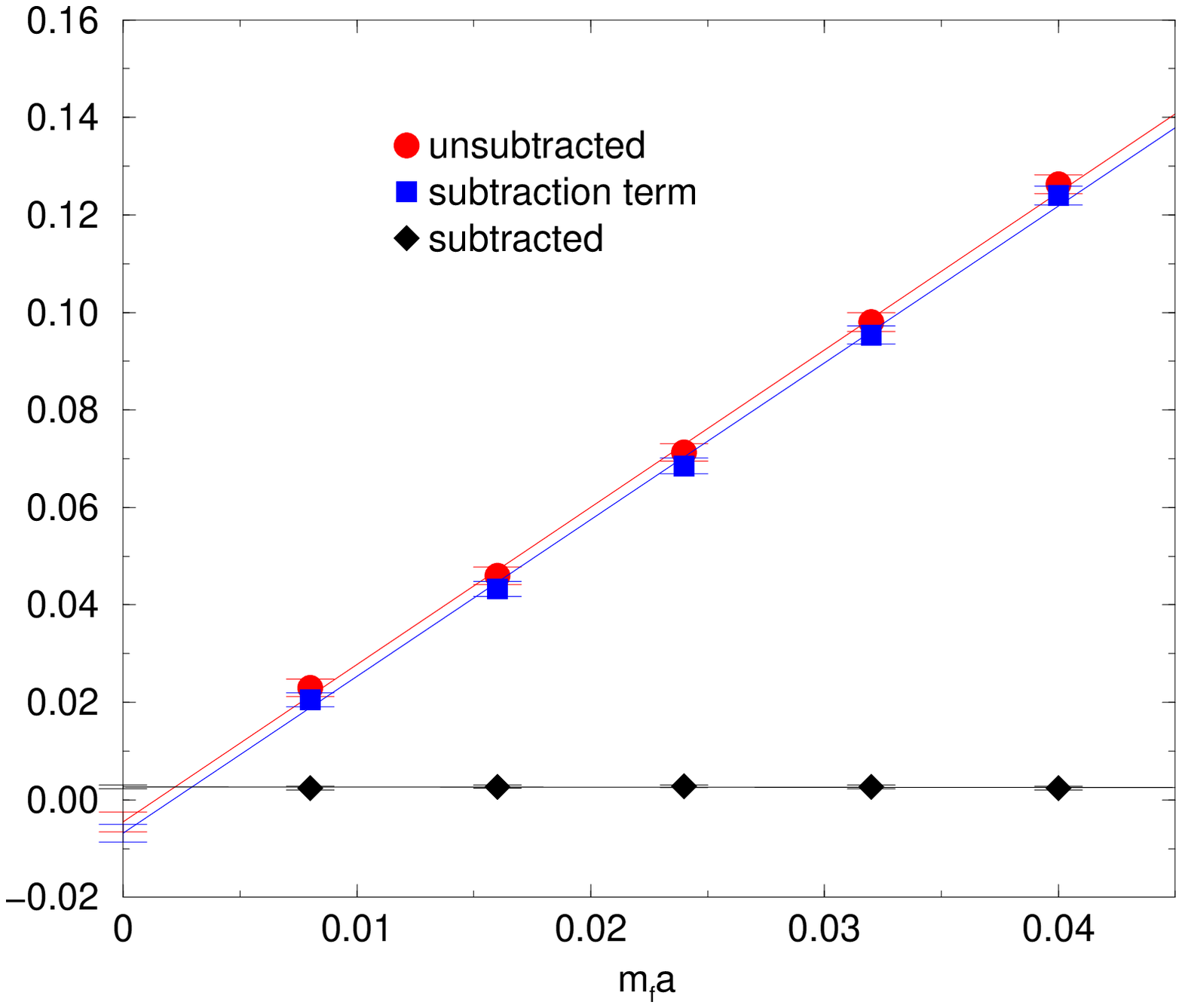}
\end{minipage}
\begin{minipage}{0.48\linewidth}
  \includegraphics[height=.28\textheight]{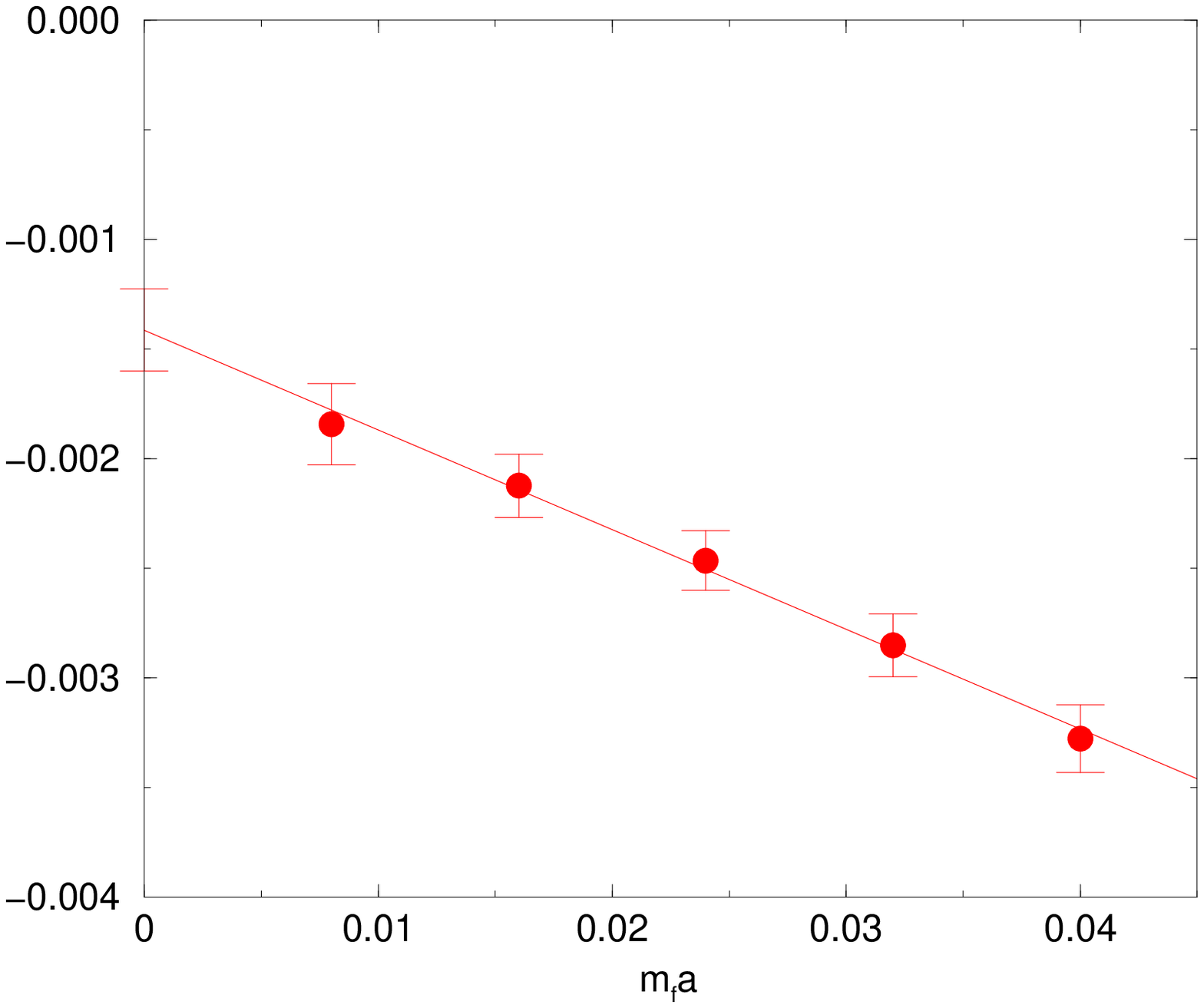}
\end{minipage}
  \caption{$K\to\pi$ matrix element of $Q^{(0)}_6$ (left) and
 $Q^{(2)}_8$ (right) as a function of $m_fa$ from Numerical Simulation I. 
In the left panel, data for matrix elements before (circle) and after
 (diamond) the subtraction and the subtraction term 
$-\alpha_6\VEV{\pi}{Q_{\rm sub}}{K}$ (square) are plotted
from 50 statistics. Linear extrapolation was used for all of plots.}
\label{KP-I}
\end{figure}

At the lowest order of chiral perturbation theory, $K\to\pi\pi$ matrix 
elements are in proportion to $K\to\pi$ matrix elements calculated on
the lattice. For $i=1$ -- $6, 9, 10$, these matrix elements are related 
as, 
\begin{eqnarray}
\VEV{\pi^+\pi^-}{Q_i^{(I)}}{K^0}&=&\frac{m_K^2-m_\pi^2}
{\sqrt{2}f}\Biggl[\frac{1}{m_{\rm PS}^2}
\VEV{\pi^+}{Q_i^{(I)}}{K^+}\Bigg|_{\rm (subt)}+{\cal O}(p^2)\Biggr],
\end{eqnarray}
in particular, for $\Delta I = 1/2$, or $I =0$, subtraction
of a lower dimension operator is needed:
\begin{eqnarray}
\VEV{\pi^+}{Q_i^{(0)}}{K^+}\Bigg|_{\rm subt} &=& 
\VEV{\pi^+}{Q_i^{(0)}-\alpha_i Q_{\rm sub}}{K^+},\label{subtraction}\\
Q_{\rm sub}= (m_s+m_d)\bar{s}d&-&(m_s-m_d)\bar{s}\gamma_5d,\ \
\alpha_i= \VEV{0}{Q_i^{(0)}}{K^0}/\VEV{0}{Q_{\rm sub}}{K^0} 
\end{eqnarray}
For $i=7,8$, we have the simpler relation
\begin{eqnarray}
\VEV{\pi^+\pi^-}{Q_i^{(I)}}{K^0}&=&-\frac{1}{\sqrt{2}f}
\VEV{\pi^+}{Q_i^{(I)}}{K^+}+{\cal O}(p^2).
\end{eqnarray}
In particular, $Q^{(0)}_6$ and $Q^{(2)}_8$ have the largest contribution
to $\epsilon'/\epsilon$, numerically. 
In FIG.~\ref{KP-I}, results of $K\to\pi$ matrix elements of these
operators are plotted. In particular, the left panel, which is the
example with $m_ca = 0.40$, shows that there is a severe cancellation in 
the subtraction in (\ref{subtraction}) for $Q^{(0)}_6$.  
Since the slope of the subtracted matrix elements have the 
error of $\sim 200\%$, we cannot quote a result of $K\to\pi\pi$ matrix 
element with the current statistics. And its depencence on $m_ca$ is not
visible, so far.

\begin{figure}
  \includegraphics[height=.28\textheight]{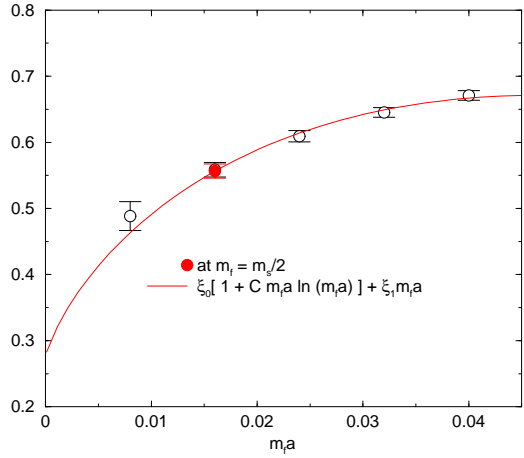}
  \caption{Lattice value of $B_K$ as a function of $m_fa$ from 77 
configurations.}
\label{BK3G}
\end{figure}

Lattice value of kaon B parameter which is defined by 
\begin{eqnarray}
B_K= \frac{\VEV{\overline{K}}{Q_{\Delta S=2}}{K}}
{8/3\VEV{\overline{K}}{A_\mu}{0}\VEV{0}{A_\mu}{K}},
\end{eqnarray}
is plotted in FIG.~\ref{BK3G} as a function of $m_fa$.
In this figure, the fit function used is 
$B_K= \xi_0[1+Cm_fa \ln (m_fa)] +\xi_1 m_fa$ with $C$ taken from
analytic result~\cite{Sharpe}. 
The physical result for $B_K$ can be obtained at $m_f = m_s/2$ 
(the filled symbol). To obtain the physical value, we are now 
calculating the $Z$ factor for $B_K$ by non-perturbative 
renormalization proposed in \cite{NPR} and the preliminary result is 
roughly consistent with the previous works~\cite{CPPACSBK,RBCep}. 
\vspace*{-0.3cm}

\section{NUMERICAL SIMULATION II }

\begin{figure}
\hspace{-0.5cm}
\begin{minipage}{0.48\linewidth}
  \includegraphics[height=.28\textheight]{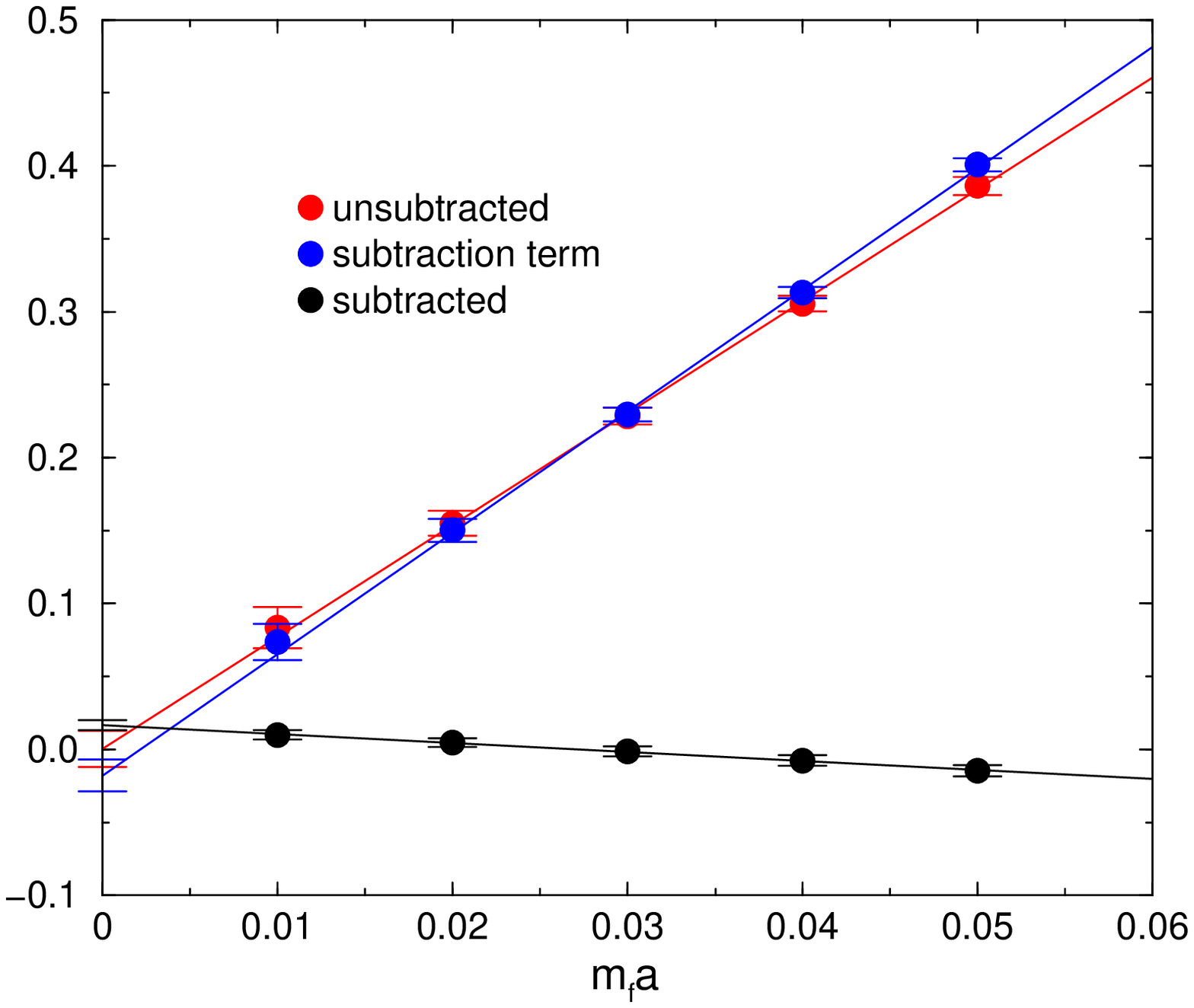}
\end{minipage}
\begin{minipage}{0.48\linewidth}
  \includegraphics[height=.28\textheight]{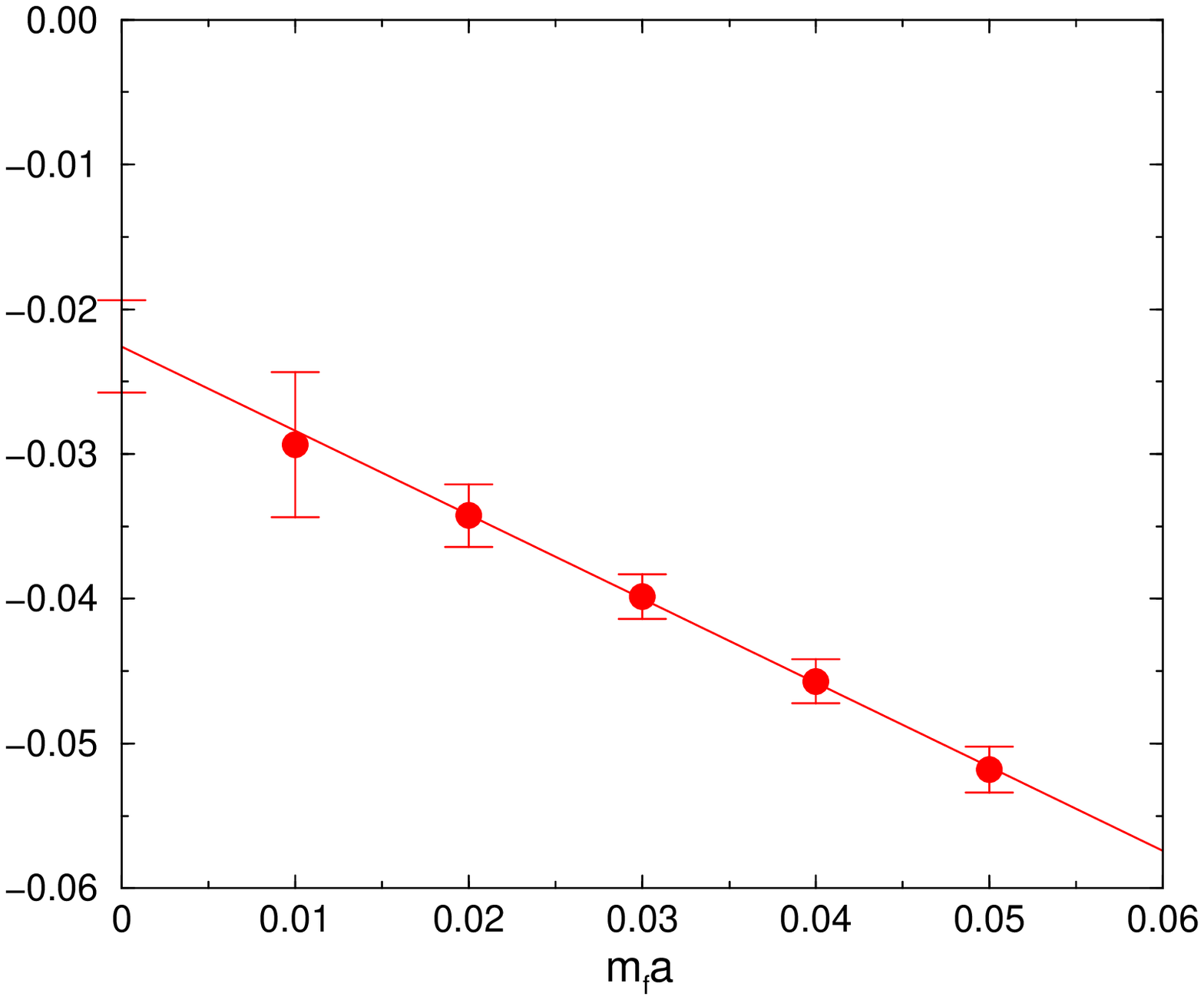}
\end{minipage}
  \caption{Same as FIG.~\ref{KP-I} but from Numerical Simulation II
with $m_{\rm sea}a= 0.03$. 72 configurations were used.}
\label{KP-II}
\end{figure}

Since the dynamical simulation demands much more resources than a quenched
one, dynamical domain-wall QCD has been explored by only our 
collaboration~\cite{RBCdyn}, so far. In this calculation, we generated 
three kinds of 
gauge configuration on a $16^3\times 32$ lattice with the mass of the sea 
quark ($u$ or $d$ quark) being
 $m_{\rm sea}a =0.02, 0.03$ and $0.04$. For each series of configurations,
$K\to\pi$ and $K\to 0$ matrix elements are calculated in the same way 
 as Simulation I with the five valence quark masses 
$m_{\rm val} = 0.01 $--$0.05$, and basic parameters 
$a^{-1}\approx 1.8$ GeV and $m_{\rm res}\approx 3$ MeV are obtained.
FIG.~\ref{KP-II} shows the same matrix elements as in FIG.~\ref{KP-I} 
as an example of the case of $m_{\rm sea}=0.03$. Although the signal seems
to be reasonable, we need much more statistics to take correct chiral limit 
$m_{\rm sea}a = m_{\rm val}\to 0$ using three data points with 
$m_{\rm sea}a = m_{\rm val}=0.02,0.03$ and $0.04$.
$B_K$ at the physical point such that $m_{\rm sea}= m_{u/d}$ and 
$m_s\sim 120$ MeV will be also obtained after a careful treatment of our 
data~\cite{GoltermanLeung} and non-perturbative renormalization which is 
now under calculation.
\vspace{0.3cm}

We thank RIKEN, BNL and the U.S.\ DOE for providing the facilities
essential for the completion of this work.




\IfFileExists{\jobname.bbl}{}
 {\typeout{}
  \typeout{******************************************}
  \typeout{** Please run "bibtex \jobname" to optain}
  \typeout{** the bibliography and then re-run LaTeX}
  \typeout{** twice to fix the references!}
  \typeout{******************************************}
  \typeout{}
 }

\end{document}